\documentclass{elsart}
\newcommand{\bm}[1]{\mbox{\boldmath$#1$}}
\usepackage[leqno]{amsmath}
\begin{document}
\date{September 8, 1997}
\begin{frontmatter}
\title{Finite--size scaling properties and Casimir \\
forces in an exactly solvable quantum \\
statistical--mechanical model\thanksref{Cong}}
\author[ISSP]{H. Chamati},
\author[IMBM]{D.M. Danchev} and
\author[ISSP]{N.S. Tonchev}
\address[ISSP]{Georgy Nadjakov Institute of Solid State Physics,
Bulgarian Academy of Sciences, Tzarigradsko chauss\'ee 72, 1784 Sofia,
Bulgaria}
\address[IMBM]{Institute of Mechanics, Bulgarian Academy of Sciences,
Acad. G. Bonchev St. bl. 4, 1113 Sofia, Bulgaria}
\thanks[Cong]
{\it Published in: Journal of Theoretical and Applied Mechanics,
Sofia (1998) 78--87}
\begin{abstract}
A $d$--dimensional finite quantum model system confined to a general
hypercubical geometry with linear spatial size $L$ and ``temporal
size'' $1/T$ ( $T$ - temperature of the system) is considered in the
spherical approximation under periodic boundary conditions. Because
of its close relation with the system of quantum rotors it
represents an effective model for studying the low--temperature
behaviour of quantum Heisenberg antiferromagnets. Close to the
zero--temperature quantum critical point the ideas of finite--size
scaling are used for studying the critical behaviour of the model.
For a film geometry in different space dimensions $\half\sigma <d<
\threehalf\sigma$ , where $0<\sigma\leq2$ controls the long--ranginess
of the interactions, an analysis of the free energy and the Casimir
forces is given.
\end{abstract}
\end{frontmatter}
\section{Introduction and brief overview of the problem}
\label{Intr}
According to the present understanding the Casimir effect is a
phenomenon common to all systems characterized by fluctuating
quantities on which some external boundary conditions are
imposed~\cite{Krech94,Most97}. Casimir forces arise from the
influence of one portion of the system on the fluctuations that
occur in a portion some distance away.

The Casimir force in statistical--mechanical systems is usually
characterized by the excess free energy due to the {\it finite--size
contributions} to the free energy of the system. The simplest and
the most often case of interest is that one of a film geometry.

A simple ${\mathcal O}(n)$ symmetric $n\geq1$ system with a geometry
$L\times\infty^2$ (and under given boundary conditions $\tau$ imposed
across the direction $L$) is a standard statistical mechanical model
for describing a fluid, or a magnet, confined between two parallel
plates of infinite area. One important quantity which arises naturally
in the thermodynamics of these confined systems is
\begin{equation}\label{eq1}
F_{(\cdots)}^\tau(T,L)=-\frac{\partial f_\tau^{\it ex}(T,L)}{\partial L},
\end{equation}
where $f_\tau^{\it ex}(T,L)$ is the excess free energy
\begin{equation}\label{eq2}
f_\tau^{\it ex}(T,L)=f_\tau(T,L)-Lf_{\it bulk}(T)
\end{equation}
Here $f_\tau(T,L)$ is the full free energy per unit area (and per
$k_BT$) of such a system and $f_{\it bulk}(T)$ is the bulk free
energy density.

In the case of a fluid (then one actually has to consider the excess
grand potential per unit area and the derivative is performed at
constant chemical potential $\mu$ and temperature $T$)
$F_{(\cdots)}$ is termed the solvation force~\cite{Evans90,Evans94}
$F_{\it (solvation)}$, (sometimes called also the disjoining
pressure) whereas in the case of a magnet one speaks instead about
the Casimir force~\cite{Krech94,Krech92} $F_{\it (Casimir)}$ (where
the derivative is performed at constant temperature and magnetic
field $H$). In the remainder we will use only the term Casimir
force.

According to the definition given by Eq.~(\ref{eq1}) the Casimir
force is a generalized force conjugated to the distance between the
surfaces bounding the system with the general property $F_{\it
(Casimir)}(T,L)\to0$, when $L\to\infty$. We will be interested in
the behaviour of $F_{\it (Casimir)}$ when $L\gg1$, which is a
condition for the applicability of the finite--size scaling theory.
In general the {\it sign} of the Casimir force is of particular
interest. It is believed that if the boundary conditions $\tau$ are
the same at the both surfaces, $F_{\it (Casimir)}$ will be {\it
negative} (see, e.g.,~\cite{Evans94,Parry92}; strictly speaking, for
an Ising--like system this should happen above the wetting
transition temperature $T_W$~\cite{Evans94,Parry92,Dietrich88}). For
boundary conditions that are not identical at both confining the
system surface planes the Casimir force is expected to be positive
~\cite{Evans94,Parry92}. For a fluid confined between walls this
implies that the net force between the plates for large separations
will be {\it attractive} in the former and {\it repulsive} in the
last case.

In general, the full free energy of a  $d$--dimensional critical
system in the form of a film with thickness $L$, area $A$, and boundary
conditions $a$ and $b$ on the two surfaces, at the bulk critical point
$T_c$, has the asymptotic form
\begin{equation}\label{eq3}
f_{a,b}(T_c,L)\cong Lf_{\it bulk}(T_c)+f^a_{\it surface}(T_c)+
f^b_{\it surface}(T_c)+L^{-(d-1)}\Delta_{a,b}+\cdots
\end{equation}
in the limits $A\to\infty$, $L\gg1$. Here $f^{(\cdot)}_{\it
surface}$ is the surface free energy contribution and $\Delta_{a,b}$
is {\it the amplitude of the Casimir interaction}. The $L$
dependence of the Casimir term (the last one in Eq.~(\ref{eq3}))
follows from the scale invariance of the free energy and has been
derived by Fisher and de Gennes~\cite{Fisher78}. The amplitude
$\Delta_{a,b}$ is {\it universal}, depending on the bulk
universality class and the universality classes of the boundary
conditions~\cite{Krech94,Krech92}.

Equation~(\ref{eq3}) is valid for both {\it fluid and magnetic
systems at criticality}. Prominent examples are, e.g.,
one--component fluid at the liquid--vapour critical point, the
binary fluid at the consolute point, and liquid $^4$He at the
$\lambda$ transition point ~\cite{Krech94}. The boundaries influence
the system to a depth given by the bulk correlation length
$\xi_\infty(T,\cdots)$ (here $\cdots$ stays for the dependence on
other essential parameters, e.g. an external magnetic field $H$). In
the vicinity of the bulk critical point
$\xi_\infty(T)\sim|T-T_c|^{-\nu}$, where $\nu$ is its critical
exponent. When $\xi_\infty(T)\ll L$ the Casimir force, as a {\it
fluctuation induced force} between the plates, is negligible. The
force becomes long--ranged when $\xi_\infty(T)$ diverges which takes
place near {\it and} below the bulk critical point $T_c$ in an
${\mathcal O}(n)$, $n\geq2$ model system in the absence of an
external magnetic field~\cite{Danchev96}. Therefore in the
statistical--mechanical systems one can turn on and off the Casimir
effect merely by changing, e.g., the temperature of the system.

The effect is of particular {\it experimental interest} in studies
of wetting of a wall by binary liquid mixtures close to their
critical end point (see, e.g.,~\cite{krech92}). Whereas the free
energy of magnetic films cannot be measured directly, the free
energy of liquid films is of relevance for surface tension
measurements and wetting phenomena. In wetting phenomena thin films
are formed such that at least one confining boundary is not rigid
but determined by thermal equilibrium ~\cite{Dietrich88,krech92}.
The Casimir force enters into the force balance and thus forms {\it
the equilibrium thickness of the wetting films}, which is accurately
measurable.

The temperature dependence of the Casimir force for two--dimensional
systems is investigated exactly only on the example of Ising
strips~\cite{Evans94}. The upper critical temperature dependence of
the force in ${\mathcal O}(n)$ models has been considered
in~\cite{Krech92}. The only example where the force is investigated
exactly as a function of both the temperature and the magnetic field
scaling variables is that of the three--dimensional spherical model
under periodic boundary conditions~\cite{Danchev96}. There exact
results for the Casimir force between two walls with a finite
separation in an $L\times\infty^2$ mean--spherical model has been
derived. The force is consistent with an {\it attraction} of the
plates confining the system . The most results available at the
moment are for the Casimir amplitudes. For $d=2$ by using
conformal--invariance methods they are exactly known for a large
class of models~\cite{Krech94}. In addition to the flat geometries
recently some results about the Casimir amplitudes between spherical
particles in a critical fluid have been derived too~\cite{Eisen95}.
For $d\neq2$ results are available via field-theoretical
renormalization group theory in $4-\varepsilon$
dimensions~\cite{Krech94,Krech92,Eisen95}, Migdal--Kadanoff
real--space renormalization group methods~\cite{Indeku}, and,
relatively recently, by Monte Carlo methods~\cite{krech96}.

In addition to the statistical mechanics the Casimir forces are
object of investigations also in the quantum electrodynamics.
Nowadays the effect is also presented in studies of topics like
nonplanar geometries of conducting or dielectric macroscopic bodies
immersed in fluids, the bag model for the description of quark
confinement in hadrons according to the quantum chromodynamics,
Casimir effect in general curved space-time in cosmology, models of
the early Universe, etc. For a review on the Casimir effect in the
aforementioned fields one can consult, e.g.,~\cite{Most97}. A
relatively recent review on the Casimir effect in
statistical--mechanical systems can be found in~\cite{Krech94}.

In recent years there has been a renewed
interest~\cite{Sachdev96,Sondhi97} in the theory of zero-temperature
quantum phase transitions. Distinctively from temperature driven
critical phenomena, these phase transitions occur at zero
temperature as a function of some non--thermal control parameter (or
a competition between different parameters describing the basic
interaction of the system), and the relevant fluctuations are of
quantum rather than thermal nature.

It is well known from the theory of critical phenomena that for the
temperature driven phase transitions quantum effects are unimportant
near critical points with $T_c>0$. It could be expected, however,
that at rather low (as compared to characteristic excitations in the
system) temperatures, the leading $T$ dependence of all observables
is specified by the properties of the zero--temperature critical
points, which take place in quantum systems. The dimensional
crossover rule asserts that the critical singularities of such a
quantum system at $T=0$ with dimensionality $d$ are {\it formally}
equivalent to those of a classical system with dimensionality $d+z$
($z$ is the dynamical critical exponent) and critical temperature
$T_c>0$. This makes it possible to investigate low--temperature
effects (considering an effective system with $d$ infinite space and
$z$ finite time dimensions) in the framework of the theory of
finite--size scaling (FSS). The idea of this theory has been applied
to explore the low--temperature regime in quantum
systems~\cite{Sachdev96,Sondhi97,Chakra89}, when the properties of
the thermodynamic observables in the finite--temperature quantum
critical region have been the main focus of interest. The most
famous model for discussing these properties is the quantum
nonlinear ${\mathcal O}(n)$ sigma model
(QNL$\sigma$M)~\cite{Sachdev96,Chakra89}.

Let us note that an increasing interest related with the {\it
spherical approximation} (or large $n$--limit) generating tractable
models in quantum critical phenomena has been observed in the last
few years ~\cite{Vojta96,Tu94,Nieu95,Nieu97,Chamati97,chamati97}.
There are different possible ways of quantization of the spherical
constraint. In general they lead to {\it different universality
classes} at the quantum critical
point~\cite{Vojta96,Tu94,Nieu95,Nieu97}.

In this paper a theory of the scaling properties of the free energy
and the Casimir forces of a quantum spherical model~\cite{Vojta96}
with nearest--neighbour and some special cases of long--range
interactions (decreasing at long distances $r$ as $1/r^{d+\sigma}$)
is presented. Only the film geometry $L\times\infty^{d-1}\times
L_\tau$ (where $L_\tau\sim\hbar/(k_BT)$ is the finite--size in the
imaginary time direction) will be considered. The plan of the paper
is as follows: we start with a brief review of the model and the
basic equations for the free energy and the quantum spherical field
in the case of periodic boundary conditions (Section~\ref{Model}).
Since we would like to exploit the ideas of the FSS theory, the bulk
system in the low--temperature region is considered like an
effective ($d+z$) dimensional classical system with $z$ finite
(temporal) dimensions. This is done to make possible a comparison
with other results based on the spherical type approximation, e.g.,
in the framework of the spherical model and the QNL$\sigma$M in the
limit $n\to\infty$. The scaling forms for the excess free energy,
the spherical field equation and the Casimir force are derived for a
$\half\sigma<d<\threehalf\sigma$ dimensional system with a film
geometry in Section~\ref{Free}. In Section~\ref{Casimir} we present
some results for the Casimir amplitudes in the case of short--range
interactions and in some special cases of long--range interactions.
The paper closes with concluding remarks given in
Section~\ref{Concl}.
\section{The model}
\label{Model}
The model we will consider here describes a magnetic ordering due to
the interaction of quantum spins. It is characterized by the
Hamiltonian~\cite{Vojta96}
\begin{equation}\label{eq4}
{\mathcal H}=\half g\sum_\ell {\mathcal P}_\ell ^2-\half\sum_{\ell\ell
^{\prime }}{\bm J}_{\ell\ell ^{\prime }}^{}{\mathcal S}_\ell ^{}{\mathcal S}
_{\ell ^{\prime }}^{}+\half\mu\sum_\ell {\mathcal S}_\ell ^2-H\sum_\ell {\mathcal
S}_\ell ^{},
\end{equation}
where ${\mathcal S}_\ell$ are spin operators at site $\ell$. The
operators ${\mathcal P}_\ell$ play the role of ``conjugated''
momenta (i.e. $[{\mathcal S}_\ell ^{}, {\mathcal S}_{\ell ^{\prime
}}^{}]=0$, $[{\mathcal P}_\ell ^{},{\mathcal P}_{\ell ^{\prime
}}^{}]=0$, and $[{\mathcal P}_\ell ^{},{\mathcal S}_{\ell ^{\prime
}}^{}]=i\delta _{\ell\ell^{\prime }}^{}$, with $\hbar =1$). The
coupling constant $g$ measures the strength of the quantum
fluctuations (below it will be called quantum parameter), $H$ is an
ordering magnetic field, and the spherical field $\mu$ is introduced
so as to ensure the constraint
\begin{equation}\label{eq5}
\sum_\ell\left<{\mathcal S}_\ell ^2\right> =N.
\end{equation}
Here $N$ is the total number of the quantum spins located at sites
``$\ell$'' of a finite hypercubical lattice $\Lambda$ of size
$L_1\times L_2\times\cdots\times L_d=N$ and $\left<\cdots\right>$
denotes the standard thermodynamic average taken with the
Hamiltonian $\mathcal H$. In~(\ref{eq4}) the coupling constants
${\bm J}_{\ell\ell ^{\prime }}^{}$ are decreasing at large distances
$|\ell-\ell^{\prime}|$ as $1/|\ell-
\ell^{\prime}|^{d+\sigma}$, where $\sigma$ determines the range of
the interaction: {\bf i)} $0<\sigma<2$ for long--range interaction
and {\bf ii)} $\sigma\geq2$ for short--range interaction.

The free energy of the model in a finite region $\Lambda$ under
periodic boundary conditions applied across the finite dimensions
has the form~\cite{chamati97}
\begin{equation}\label{eq6}
\beta f_\Lambda\left(\beta ,g,H\right) =\sup_\mu\left\{\frac 1N
\sum_{q}\ln\left[ 2\sinh\left(\half\beta\omega
\left( {\bm q};\mu\right)\right)\right] -\frac{\mu}{2}\beta
-\frac{\beta g{\bm J}}
{2\omega ^2\left(0;\mu\right) }H^2\right\} .
\end{equation}
Here the vector $\bm q$ is a collective symbol, which for $L_j$ odd
integers has the components $\left\{\frac{2\pi n_1}{L_1},\cdots ,
\frac{2\pi n_d}{L_d}\right\} , n_j\in\left\{ -\frac{L_j-1}2,\cdots ,
\frac{L_j-1}2\right\}$, and $\beta$ is the inverse temperature with
the Boltzman constant $k_B=1$. In~(\ref{eq6}) the spectrum is
$\omega^2\left( {\bm q};\mu\right) =g\left(\mu + U({\bm q})\right)$
with $U({\bm q})=2{\bm J}\sum_{i=1}^d\left(1-\cos q_i\right)$ for
nearest neighbour interactions and will be taken of the form $U({\bm
q})\cong{\bm J}\rho_\sigma|{\bm q}|^\sigma$, $0<\sigma<2$, for long
range interactions ($\rho_\sigma>0$ is a parameter to be taken equal
to one in the remainder). In the above expressions $U({\bm q})$ is
the Fourier transform of the interaction matrix where the energy
scale has been fixed so that $U(0)=0$. The supremum in
Eq.~(\ref{eq6}) is attained at the solutions of the mean--spherical
constraint, Eq.~(\ref{eq5}), that reads
\begin{equation}\label{eq7}
1=\frac tN\sum_{m=-\infty }^\infty\sum_{q}\frac 1{\phi
+U({\bm q})/{\bm J}+b^2m^2}+\frac{h^2}{\phi ^2},
\end{equation}
where we have introduced the notations: $b=(2\pi t)/\lambda$,
$\lambda=\sqrt{g/\bm J}$ is the normalized quantum parameter,
$t=T/{\bm J}$ - the normalized temperature, $h=H/\sqrt{\bm J}$ - the
normalized magnetic field, and $\phi=\mu/{\bm J}$ is the scaled
spherical field. Eqs.~(\ref{eq6}) and~(\ref{eq7}) provide the basis
of the study of the critical behaviour of the model under
consideration.

In the thermodynamic limit it has been shown~\cite{Vojta96} that for
$d>\sigma$ the long--range order exists at finite temperatures up to
a given critical temperature $t_c(\lambda)$. Here we shall consider
the {\it low--temperature region} for
$\half\sigma<d<\threehalf\sigma$. We remind that $\half\sigma$ and
$\threehalf\sigma$ are the lower and the upper critical dimensions,
respectively, for the quantum critical point of the considered
system.
\section{Scaling form of the excess free energy and the Casimir force at
low temperatures}
\label{Free}
For a system with a film geometry  $L\times\infty^{d-1}\times L_\tau$
(where $\half\sigma<d<\threehalf\sigma$), after taking the limits
$L_2\to\infty,\cdots,L_d\to\infty$ in Eq.~(\ref{eq6}) and by using the
Poisson summation formula for the only remaining finite space
dimensionality $L_1=L$, we receive the following expression for the
full free energy density
\begin{eqnarray}\label{eq8}
f(t,\lambda,h;L)&=&-\frac{\lambda}{4\sigma}\frac{k_d}{\sqrt{\pi}}
\Gamma\left(\frac{d}{\sigma}\right)\Gamma\left(-\frac{d}{\sigma}-
\frac{1}{2}\right)\phi^{\frac{d}{\sigma}+\frac{1}{2}}+
\left(\frac{\lambda}{\lambda_c}-1\right)\frac{\phi}{2}-\frac{h^2}{2\phi}\\
& &-\frac{\lambda}{\sigma}\frac{k_d}{\sqrt{\pi}}
\Gamma\left(\frac{d}{\sigma}\right)\phi^{\frac{d}{\sigma}+\frac{1}{2}}
\sum_{m=1}^{\infty}K_{\frac{d}{\sigma}+\frac{1}{2}}
\left(m\frac{\lambda}{t}\phi^\frac{1}{2}\right)
\left(m\frac{\lambda}{2t}\phi^\frac{1}{2}\right)^{-\left(\frac{d}{\sigma}
+\frac{1}{2}\right)}\nonumber\\
& &+2\frac{tL^{-\frac{d-2}{2}}}{(2\pi)^\frac{d}{2}}\!\sum_{m=1}^{\infty}
\!\int\limits_0^{x_D}\frac{dx}{m^\frac{d-2}{2}}x^\frac{d}{2}J_{\frac{d}{2}-1}(mLx)
\!\ln\!\!\left[\!2{\rm sinh}\!\left(\frac{\lambda}{2t}\sqrt{\phi+x^\sigma}\right)
\!\!\right]\!,\nonumber
\end{eqnarray}
where $k_d^{-1}=\half(4\pi)^\frac{d}{2}\Gamma(d/2)$, $x_D$ is the radius
of the sphericalized Brillouin zone, $K_\nu(x)$ and
$J_\nu(x)$ are the MacDonald and Bessel functions, respectively, and the
critical value of $\lambda=\lambda_c$ is
\begin{equation}\label{eq9}
\lambda_c^{-1}=\half(2\pi)^{-d}\int\limits d^d{\bm q}(U({\bm q})/{\bm J})
^{-\frac{1}{2}}.
\end{equation}
In Eq.~(\ref{eq8}) $\phi$ is the solution of the corresponding
spherical field equation that follows by requiring the partial
derivative of the r.h.s. of Eq.~(\ref{eq8}) with respect to $\phi$
to be zero. The bulk free energy $f_{\it bulk}(t,\lambda,h)$ results
from $f(t,\lambda,h;L)$ by merely taking the limit $L\to\infty$ in
it. Let us denote the solution of the corresponding bulk spherical
field equation by $\phi_\infty$. Then for the excess free energy it
is possible to obtain the finite size scaling form
\begin{equation}\label{eq10}
f^{\it ex}(t,\lambda,h;L)/L=\lambda L^{-(d+z)}X(x_1,x_2,a),
\end{equation}
with scaling variables
$x_1=L^{-1/\nu}\left(1/\lambda-1/\lambda_c\right)$,
$x_2=hL^{\Delta/\nu}$ and $a=tL^z/\lambda$. Here
$\nu^{-1}=d-\half\sigma$,
$\Delta/\nu=\half\left(d+\threehalf\sigma\right)$ and
$z=\half\sigma$ are the critical exponents of the
model~\cite{Vojta96}. In Eq.~(\ref{eq10}) the {\it universal}
scaling function of the excess free energy may obtained in an
explicit form (this will be presented in a subsequent article).

For the Casimir forces in the considered system
\begin{equation}\label{eq12}
F_{\it Casimir}(T,\lambda,h;L)=\lambda L^{-(d+z)}X_{\it Casimir}(x_1,x_2,a),
\end{equation}
where the {\it universal} scaling functions of the Casimir force
$X_{\it Casimir}(x_1,x_2,a)$ is related to that one of the excess
free energy $X\equiv X(x_1,x_2,a)$ by
\begin{equation}\label{eq13}
X_{\it Casimir}(x_1,x_2,a)=-(d+z)X-\frac{1}{\nu}x_1\frac{\partial X}{\partial x_1}
+\frac{\Delta}{\nu}x_2\frac{\partial X}{\partial x_2}
+za\frac{\partial X}{\partial a}.
\end{equation}
\section{Casimir amplitudes}
\label{Casimir}
In this section we determine the Casimir amplitudes of the model for
the case of short range interactions at $d=2$ and for some special
cases of long range interactions.
\subsection{short range interactions ($d=2$)}
It can be shown that the solution $y_0$ of the spherical field
equation for the finite system with a film geometry
$L\times\infty\times L_\tau$ at zero temperature (i.e. $1\ll
L\ll\infty$, $L_\tau=\infty$) is
$y_0=\ln\left(\sqrt{5}/2+1/2\right)$ at the quantum critical point
$\lambda=\lambda_c$, $h=0$ (at this point
$y_\infty=0$)~\cite{Chamati97,chamati97}. Setting this value of
$y_0$ in the scaling function~(\ref{eq10}) of the excess free
energy, taking into account that
$K_{\frac{3}{2}}=\sqrt{\pi/(2x)}\exp(-x)(1+1/x)$, the identity
$\ln\left[2{\rm sinh}\left(\sqrt{y}/2\right)\right]=\sqrt{y}/2 -
{\rm Li}_1\left[\exp\left(-\sqrt{y}\right)\right]$ and the
properties of the polylogarithm functions ${\rm
Li}_p(x)$~\cite{Sachdev93}, we obtain from Eqs.~(\ref{eq3})
and~(\ref{eq10}) that the Casimir amplitude is
\begin{equation}\label{eq14}
\Delta_{\it periodic\ b.c.}=-\frac{2\zeta(3)}{5\pi}\approx-0.153051.
\end{equation}
Here $\zeta(3)$ is the Riemann zeta function.
\subsection{long range interactions ($d=\sigma$)}
Let us consider the "temporal Casimir amplitude" in a system with a
geometry $\infty^d\times L_\tau$ with $d/\sigma=1$ at the quantum
critical point $\lambda=\lambda_c$, $h=0$. In a way, similar to that
one explained for the case considered previously one obtains
($0<\sigma\leq2$)
\begin{equation}\label{eq15}
f(t,\lambda_c,0;\infty)-f(0,\lambda_c,0;\infty)=-\lambda_c
\frac{16}{5\sigma}\frac{\zeta(3)}{(4\pi)^{\sigma/2}}
\frac{t^3}{\Gamma(\sigma/2)}.
\end{equation}
From here one can identify the "temporal Casimir amplitude" to be
\begin{equation}\label{eq16}
\Delta_t(\sigma)=-
\frac{16}{5\sigma}\frac{\zeta(3)}{(4\pi)^{\sigma/2}}
\frac{1}{\Gamma(\sigma/2)}.
\end{equation}

\section{Concluding remarks}
\label{Concl}
In the present article the free energy of a system with a geometry
$L\times\infty^{d-1}\times L_\tau$ (where
$\half\sigma<d<\threehalf\sigma$), is derived (see Eq.~(\ref{eq8})).
For $\sigma=2$ this new result reduces to the one reported
in~\cite{chamati97} where only the case of short--range interactions
has been considered. A general expression for the Casimir force in
the {\it quantum} spherical model is obtained (see
Eqs.~(\ref{eq12},\ref{eq13}). In the classical limit ($\lambda=0$)
for a system with short--range interaction it coincides with the
corresponding one derived in~\cite{Danchev96} for the classical
spherical model. Except for some two--dimensional systems the only
model within which the Casimir effect is investigated in an exact
manner for $d\neq2$ is that one of the spherical model. The
investigations presented here complement the aforementioned ones for
the case when quantum fluctuations are present in the system. In
order to derive in a simple closed form the Casimir amplitudes some
particular cases have been considered ($d=\sigma$). For $d=2$ this
amplitude is given in Eq.~(\ref{eq14}). This amplitude is equal to
the "temporal Casimir amplitude" (i.e. the corresponding temperature
corrections in an $\infty^2\times L_\tau$ system to the ground state
of the bulk system) for the ${\mathcal O}(n)$ sigma model in the
limit $n\to\infty$~\cite{Sachdev93}. We have demonstrated here {\it
explicitly} that the two models, due to the fact that they belong to
the same universality class, indeed possess equal Casimir amplitudes
as it is to be expected on the basis of the {\it hypothesis of the
universality}. We note that instead of considering a finite system
with a film geometry at the zero temperature one can consider a bulk
system at low temperatures. As it is already clear from above this
leads to the same result f or the Casimir amplitudes. Note, that in
accordance with the general expectations these amplitudes are {\it
negative}. The correction to the ground state energy of the bulk
system due to the nonzero temperature is determined for the general
case $d=\sigma$ by Eq.~(\ref{eq16}). Note, that the defined there
"temporal Casimir amplitude" $\Delta_t(\sigma)$ reduces for
$\sigma=2$ to the "normal" Casimir amplitude, given by
Eq.~(\ref{eq14}). This reflects the existence of a special symmetry
for that case between the "temporal" and the space dimensionalities
of the system.

\section*{Acknowledgements}
This \ work \ is supported by \ The Bulgarian \ Science Foundation
(Projects F608/96 and MM603/96).

\end{document}